\documentclass[aps,prb,reprint,superscriptaddress,noeprint,showkeys]{revtex4-2}

\usepackage[hidelinks]{hyperref}
\usepackage[dvipsnames]{xcolor}
\hypersetup{
    colorlinks,
    linkcolor={blue},
    citecolor={blue},
    urlcolor={blue},
}
\usepackage{graphicx}
\usepackage{amsmath,braket,graphicx,mathtools,amssymb}
\usepackage{amsfonts}
\newcommand{\CrCl}{$\mathrm{CrCl_3}$}
\newcommand{\CrBr}{$\mathrm{CrBr_3}$}
\newcommand{\CrI}{$\mathrm{CrI_3}$}
\newcommand{\CrX}{$\mathrm{CrX_3}$}

\begin{document}


\title{Enhanced Magnetization by Defect-Assisted Exciton Recombination\\ in Atomically Thin $\mathrm{CrCl_3}$}

\author{Xin-Yue Zhang}
\thanks{These authors contributed equally to this work.}
\author{Thomas K. M. Graham}
\thanks{These authors contributed equally to this work.}
\affiliation{Department of Physics, Boston College, Chestnut Hill, MA, 02467, USA}
\author{Hyeonhu Bae}
\thanks{These authors contributed equally to this work.}
\affiliation{Department of Condensed Matter Physics, Weizmann Institute of Science, Rehovot, Israel}
\author{Yu-Xuan Wang}
\affiliation{Department of Physics, Boston College, Chestnut Hill, MA, 02467, USA}
\author{Nazar Delegan}
\affiliation{Materials Science Division, Argonne National Laboratory, Lemont, IL, 60439, USA}
\affiliation{Center for Molecular Engineering, Argonne National Laboratory, Lemont, IL, 60439, USA}
\affiliation{Pritzker School of Molecular Engineering, University of Chicago, Chicago, IL 60637, USA}
\author{Jonghoon Ahn}
\affiliation{Materials Science Division, Argonne National Laboratory, Lemont, IL, 60439, USA}
\affiliation{Center for Molecular Engineering, Argonne National Laboratory, Lemont, IL, 60439, USA}
\author{Zhi-Cheng Wang}
\affiliation{Department of Physics, Boston College, Chestnut Hill, MA, 02467, USA}
\author{Jakub Regner}
\affiliation{Department of Inorganic Chemistry, University of Chemistry and Technology Prague, Technicka 5, 166 28, Prague 6, Czech Republic}
\author{Kenji Watanabe}
\affiliation{Research Center for Electronic and Optical Materials, National Institute for Materials Science, 1-1 Namiki, Tsukuba 305-0044, Japan}
\author{Takashi Taniguchi}
\affiliation{Research Center for Materials Nanoarchitectonics, National Institute for Materials Science,  1-1 Namiki, Tsukuba 305-0044, Japan}
\author{Minkyung Jung}
\affiliation{DGIST Research Institute, DGIST, Daegu 42988, Republic of Korea}
\author{Zden\v ek Sofer}
\affiliation{Department of Inorganic Chemistry, University of Chemistry and Technology Prague, Technicka 5, 166 28, Prague 6, Czech Republic}
\author{Fazel Tafti}
\affiliation{Department of Physics, Boston College, Chestnut Hill, MA, 02467, USA}
\author{David D. Awschalom}
\author{F. Joseph Heremans}
\affiliation{Materials Science Division, Argonne National Laboratory, Lemont, IL, 60439, USA}
\affiliation{Center for Molecular Engineering, Argonne National Laboratory, Lemont, IL, 60439, USA}
\affiliation{Pritzker School of Molecular Engineering, University of Chicago, Chicago, IL 60637, USA}
\author{Binghai Yan}
\affiliation{Department of Condensed Matter Physics, Weizmann Institute of Science, Rehovot, Israel}
\author{Brian B. Zhou}
\email{brian.zhou@bc.edu}
\affiliation{Department of Physics, Boston College, Chestnut Hill, MA, 02467, USA}

\date{\today}

\begin{abstract}
Two dimensional (2D) semiconductors present unique opportunities to intertwine optical and magnetic functionalities and to tune these performances through defects and dopants. Here, we integrate exciton pumping into a quantum sensing protocol on nitrogen-vacancy centers in diamond to image the optically-induced transient stray fields in few-layer, antiferromagnetic \CrCl. We discover that exciton recombination enhances the in-plane magnetization of the \CrCl{} layers, with a predominant effect in the surface monolayers. Concomitantly, time-resolved photoluminescence measurements reveal that nonradiative exciton recombination intensifies in atomically thin \CrCl{} with tightly localized, nearly dipole-forbidden excitons and amplified surface-to-volume ratio. Supported by experiments under controlled surface exposure and density functional theory calculations, we interpret the magnetically enhanced state to result from a defect-assisted Auger recombination that optically activates electron transfer between water vapor related surface impurities and the spin-polarized conduction band. Our work validates defect engineering as a route to enhance intrinsic magnetism in single magnetic layers and opens a novel experimental platform for studying optically-induced, transient magnetism in condensed matter systems.




\end{abstract}

\maketitle

Magnetic order in 2D semiconductors can modulate their reflection \cite{Huang2017,Gong2017a} or absorption \cite{Zhang2019,Grzeszczyk2023} of polarized light, as well as the polarization \cite{Seyler2018,Wang2021} or spectrum \cite{Wilson2021} of their photoluminescence (PL). These magneto-optical effects, which allow sensitive optical probing of the underlying magnetization, often hinge on the special properties of bound electron-hole pairs, known as excitons, that exhibit the enhanced Coulomb interactions \cite{Wu2019a,Molina-Sanchez2020}, spin-valley coupling \cite{Choi2023}, or interlayer hybridization \cite{Wilson2021} of the 2D limit. An intriguing prospect is whether the inverse effect occurs: can excitonic processes in a 2D semiconductor modify its magnetic properties? Such optical control of magnetism could enable low-power, high-speed spintronic devices featuring the versatility of the 2D materials platform \cite{OrtizJimenez2021,Zhang2022,Wang2022,Hao2022,Dabrowski2022,Diederich2022,Bae2022,Qiu2023a}.

In 2D magnets, such as the chromium trihalides (\CrX, X = Cl, Br, I) \cite{Wu2019a,Molina-Sanchez2020}, transition metal dihalides MX$_2$ (X = Cl, Br, I) \cite{McGuire2017a,Occhialini2024}, and transition metal phosphorus trichalcogenides MPX$_3$ (X = S, Se) \cite{Kang2020,Birowska2021a,Belvin2021,Afanasiev2021}, bonding between the transition metal and ligand atoms exhibits more ionic character than in the transition metal dichalcogenides (TMDs). Ligand field theory thus serves as a starting point for understanding the optical excitations of 2D magnets, although considerations from band theory become necessary as the covalency of bonding increases \cite{Wu2019a,Molina-Sanchez2020,Acharya2022a}. In 2D magnets, excitons, appearing as sharp sub-bandgap peaks in the optical absorption, can be traced to the localized $d$\mbox{-}$d$ transitions on the transition metal ions. Although parity-forbidden optically in the isolated atomic limit, these transitions become weakly allowed in crystals due to metal-ligand ($dp$) hybridization, as well as due to further symmetry breaking caused by the spin-orbit interaction or coupling to phonons and defects \cite{Acharya2022a}. Consequently, these excitons can possess near-atomic spatial confinement and strong binding energies \cite{Wu2022,Zhu2020b} characteristic of Frenkel excitons in ionic insulators \cite{Abbamonte2008} or molecular crystals \cite{Lettmann2021}, exceeding that of Wannier-Mott excitons in the TMDs \cite{Wang2018c}.

The strongly-correlated nature of excitons plays a vital role in imprinting the magnetic state of 2D semiconductors onto their optical response, for example through giant Kerr and Faraday effects \cite{Huang2017,Wu2019a}, spin-dependent absorption \cite{Zhang2019,Grzeszczyk2023} and photoluminescence \cite{Seyler2018,Wang2021,Wilson2021}, or ultra-narrow exciton linewidths stabilized by spin-orbit coherence \cite{Kang2020,He2024}. Alternatively, from the point of view of optical control, ultrafast pump-probe measurements have demonstrated that $d$\mbox{-}$d$ excitons can generate magnons in bulk samples of NiPS$_3$ \cite{Belvin2021,Afanasiev2021}, while circularly polarized ultrafast pulses at the charge-transfer exciton energy were shown to switch the magnetic state of three-layer CrI$_3$ \cite{Zhang2022}. Nevertheless, the distinct interactions of these strongly-bound excitons with defects and interfaces, and the novel routes by which they can modulate magnetism, particularly in atomically-thin samples, have yet to be fully elucidated.

In this work, we introduce pump-probe nitrogen-vacancy (NV) center magnetometry for the spatiotemporal imaging of optically controlled magnetism. By pumping the excitonic resonances in few-layer, antiferromagnetic (AF) \CrCl, we discover that its in-plane, layer-alternating magnetization is remarkably increased within the pump spot by up to 0.5~$\mu_B/\mathrm{nm}^2$ in odd-layer regions, where $\mu_B$ is the Bohr magneton. The transient stray field due to this optical enhancement provides sensitive even-odd layer contrast as a local signature in the flake's interior, in distinction to the static stray field that diminishes away from the edges.

We correlate the formation of the magnetically enhanced state with a defect-assisted Auger recombination \cite{Chen2001, Wang2015g} that predominates for excitons in \CrCl{}, the most Frenkel-like within the \CrX{} family \cite{Acharya2022a}. Time-resolved PL measurements probing the dynamics of exciton recombination reveal that this nonradiative pathway intensifies in thinner, unencapsulated flakes, confirming the role of surface impurities. Crucially, regions with higher rates of nonradiative exciton recombination display stronger optically-induced magnetization. The latter persists for tens of microseconds, exceeding the exciton lifetime and reflecting the slow charge trapping dynamics after recombination. Through measurements on a \CrCl{} flake sequentially exposed to water vapor (H$_2$O) and oxygen gas (O$_2$), we demonstrate control over the amplitude of the transient magnetization and pinpoint H$_2$O exposure as the primary catalyst. Our density functional theory (DFT) calculations support that H$_2$O adsorbates can introduce midgap states that mediate Auger recombination of excitons. This process could enhance the magnetization of \CrCl{} by transferring unpolarized defect electrons into the majority spin conduction band of its host surface layer.



Our results provide a demonstration of defect-modulation of the magnetic properties of single atomic layers with intrinsic order. Although proposed in numerous computational studies \cite{Zhao2018,Wang2018d,Pizzochero2020,Guo2018a,Yang2021,Luo2020}, this tuning knob, with strategies unique to the 2D limit such as molecular adsorption demonstrated here, has eluded experimental realization due to stringent requirements on control samples and quantitative magnetic measurement. Our light-activated charge transfer is distinct from other mechanisms for optical control of magnetism, such as through the absorption-induced occupation of a short-lived excited state with different magnetic exchange \cite{Mikhaylovskiy2020,Ron2020} or through nonlinear field-induced effects, such as the inverse Faraday effect or inverse Cotton-Mouton effects, that require high intensity and suitably polarized light \cite{Kimel2007}. Charge transfer effects have been observed in ensembles of molecular magnets \cite{Shimamoto2002} and semiconductor quantum dots \cite{Pinchetti2018}, but are revealed here for an atomically thin antiferromagnet through a pioneering, spatially-resolved technique. Thus, our discovery opens a high-speed, high spatial resolution avenue to control the doping and magnetic properties of atomically-thin magnets using light.

\begin{figure*}[hbt]
\includegraphics[scale=1]{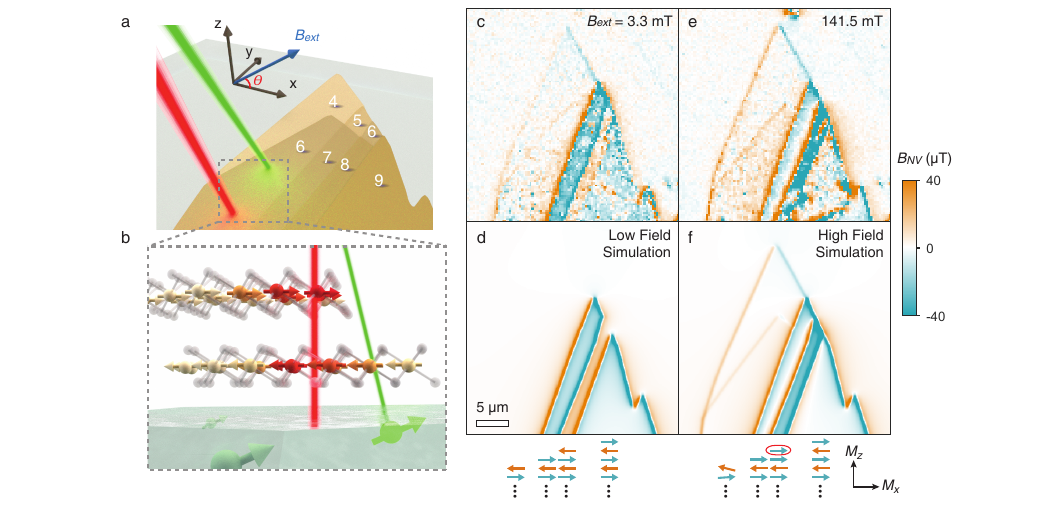}
\caption{\label{fig:1}Experimental background. (a) Few-layer \CrCl{} flakes are transferred onto a diamond substrate (gray) containing a near-surface ensemble of NV centers. NV magnetic imaging is performed by scanning the green probe beam (515 nm), while exciton dynamics in \CrCl{} are stimulated by the ``red'' pump beam (variable wavelength). (b) The layer magnetization in \CrCl{} lies in-plane and alternates antiferromagnetically between layers, producing a stray magnetic field sensed by proximal NV centers (green). (c) Experimental dc stray field ($B_{NV}$) image of a \CrCl{} flake, with 4 to 9-layer regions as labeled in panel~a), at low external field $B_{ext}$. (d) Simulation of $B_{NV}$ assuming in-plane, AF interlayer order. The bottom cartoon displays the magnetization of the top-most layers for the vertically adjacent region, showing odd-layer regions to possess an uncompensated monolayer along the in-plane projection of $B_{ext}$. (e) Experimental $B_{NV}$ image at high $B_{ext}$. (f) Simulation of $B_{NV}$ assuming that one layer of spins (circled) within the 6/8-layer stripe flips, resulting in two uncompensated layers.}
\end{figure*}

\subsection{In-Plane, AF Interlayer Order}
Below its N\'eel temperature $T_N \approx$ 17 K, bulk \CrCl{} possesses moments that order ferromagnetically in-plane within each layer, but antiferromagnetically between adjacent layers \cite{McGuire2017}. As in-plane magnetization cannot be detected via magneto-optical effects using light at normal incidence, magnetic order in atomically-thin \CrCl{} has only been investigated by tunneling magnetoresistance \cite{Cai2019a,Wang2019b,Klein2019,Kim2019c} and x-ray magnetic circular dichroism at grazing incidence \cite{Bedoya-Pinto2021}, which both possess limited spatial resolution. To overcome this lack of spatial information, we leverage magnetic imaging by NV spins in diamond, previously employed to visualize the out-of-plane magnetic order in \CrI{} and \CrBr{}  \cite{Thiel2019,Sun2020,Song2021}. In extension to prior works, here we implement a geometry optimized for high-sensitivity ac magnetometry \cite{Zhang2021} and pump-probe optical access \cite{Wang2023}. Few-layer \CrCl{} flakes, encapsulated by hBN unless otherwise described, are transferred inside an argon-filled glovebox onto a $^{12}$C isotopically-enriched diamond substrate containing a near-surface NV ensemble with prolonged coherence time (Appendix A). As illustrated in Figs. \ref{fig:1}a,b, the pump beam of variable wavelength (405-785 nm; shown in red) excites exciton dynamics in \CrCl{}, while an independently steered probe beam (green, 515~nm) images the resulting stray magnetic field across the NV ensemble at diffraction-limited spatial resolution. 

We begin by performing pulsed optically detected magnetic resonance (ODMR) to determine the static, layer-dependent magnetization in \CrCl{} flakes at $T$ = 4~K. In this mode, the NV spin transitions are Zeeman-shifted by the dc stray field along the NV center axis, angled at $\theta = 35^\circ$ from the surface in the $xz$-plane (Fig. \ref{fig:1}a) \cite{Thiel2019}. We apply a bias field $B_{ext}$ along the NV center axis and define the quantity $B_{NV}$ to be solely the field due to the flake by subtracting $B_{ext}$ from the measured total field.

Figure \ref{fig:1}c shows the image of $B_{NV}$ at small $B_{ext}$ = 3.3~mT for the \CrCl{} flake depicted in Fig. \ref{fig:1}a, containing regions with four to nine layers. Significant stray fields are detected over the odd-layer regions, in contrast to negligible stray fields over the even-layer regions, except over a narrow 6/8-layer stripe where the signal is contributed by its two neighboring odd layers. In Fig.~\ref{fig:1}d, we simulate the flake's stray field assuming AF interlayer magnetic order oriented along the in-plane projection of $B_{ext}$ ($+\hat{x}$) \cite{Fabre2020}. This simulation reproduces the salient even-odd stray field contrast, as well as the detailed spatial parity of the stray field, which is inconsistent with out-of-plane magnetic order (Supplementary Fig. 1). Quantitative linecuts presented in Supplementary Fig. 1e reveal that the areal magnetization of the 5/7-layer stripe is approximately 14 $\mu_b$/nm$^2$, slightly reduced from the expected saturation magnetization of a single uncompensated layer (19 $\mu_b$/nm$^2$).

Importantly for later discussions, we observe a reversal of the direction of the stray field over the narrow 6/8-layer stripe when $B_{ext}$ is increased to 141.5 mT (Fig. \ref{fig:1}e). Our simulation (Fig. \ref{fig:1}f) and extracted linecut (Supplementary Fig. 1i) explain that this reversal corresponds to the flipping of a single layer that is anti-aligned with the external field, such that the net magnetization of the even-layer stripe increases from approximately fully-compensated to two uncompensated layers (28 $\mu_B$/nm$^2$). For the outer even-layer regions (4 and 6-layer), the stray fields at their boundaries become stronger at $B_{ext}$ = 141.5 mT, with a weak positive field in the 6-layer's interior, signaling the emergence of finite magnetizations due to canting of the spins out of the plane. 


\begin{figure}[t]
\includegraphics[scale=1]{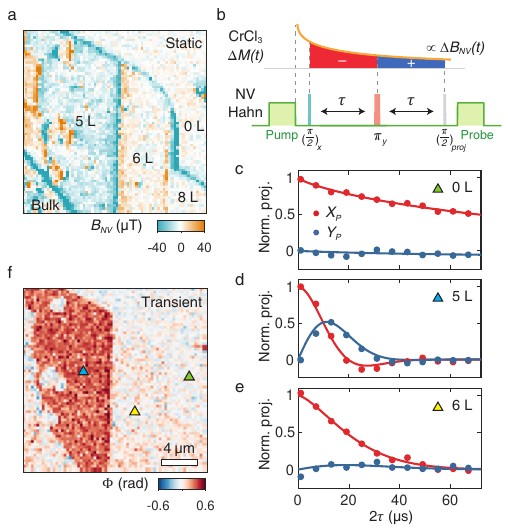}
\caption{\label{fig:2}Transient stray fields from optically-excited \CrCl. (a) Preliminary: dc stray field ($B_{NV}$) image identifying 5, 6, 8-layer and bulk regions of a \CrCl{} flake. (b) Schematic of a Hahn echo sequence with total duration $2\tau$ applied to the NV center, alongside a temporally decaying magnetic field $\Delta B_{NV}(t)$ due to a magnetization change $\Delta M(t)$ in \CrCl{} induced by the initial green pulse (power $P = 80\;\mu$W). The $\pi$-pulse negates the phase accumulated during the first half of the evolution. (c,d,e) Hahn echo measured over: c) the bare NV substrate (0-layer), d) 5-layer region, and e) 6-layer region. Only the 5-layer region engenders a non-zero NV precession angle $\Phi$, indicated by the simultaneous evolution of the $X_P$ and $Y_P$ projections of the NV superposition state with increasing Hahn echo duration The solid lines in (c-e) are fits assuming an exponential decoherence envelope and a possible linearly increasing $\Phi$ with $2\tau$ (Supplementary Section IV). (f) Image of the precession angle $\Phi$, reflecting the optically-induced transient stray field, for an Hahn echo sequence with $2\tau = 13\;\mu$s over the same region as a) by scanning the single green beam ($B_{ext} = 141.5$~mT).}
\end{figure}

\subsection{Optical Modulation of N\'eel Magnetization}
Having established the underlying in-plane, layered AF state, we now uncover a startling optical effect on the N\'eel magnetization of atomically thin \CrCl{}. Figure \ref{fig:2}a displays the dc stray field image at $B_{ext}$ = 141.5~mT for another \CrCl{} flake containing 5, 6, and 8-layer regions. Significant $B_{NV}$ is again only observed over the odd-layer area.

\begin{figure*}[hbt]
\includegraphics[scale=1]{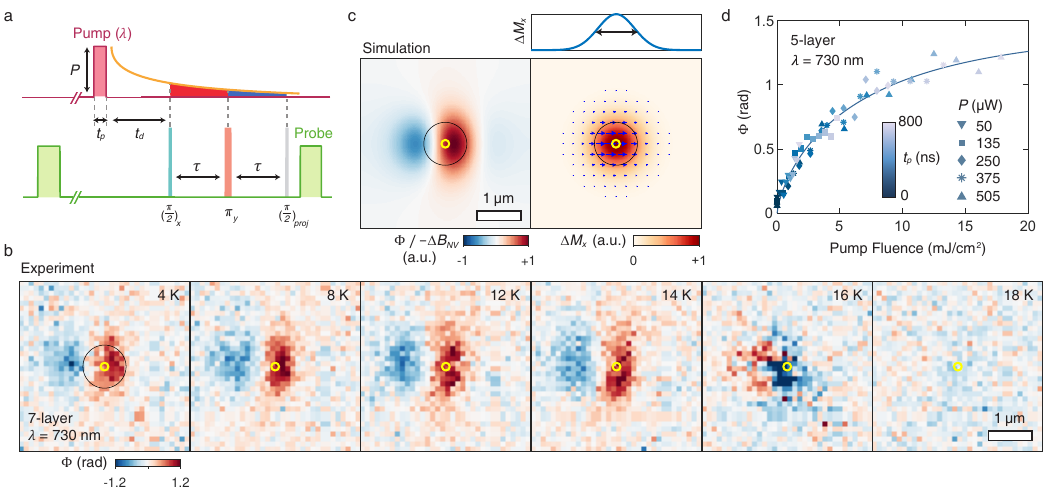}
\caption{\label{fig:3}Pump-probe imaging of exciton-enhanced magnetization. (a) Experimental timeline integrating a pump beam (wavelength $\lambda$) into the Hahn echo sequence. The pump pulse of width $t_p$ and power $P$ occurs at least $50\;\mu$s after the initial green pulse (NV initialization); the Hahn echo is commenced after a delay $t_d$ from the pump pulse. (b) Temperature evolution of the local Hahn precession angle $\Phi(\bold{r})$ on a 7-layer region, measured by scanning the probe beam relative to a fixed pump beam at the yellow circle. $\Phi(\bold{r})$ images the spatial distribution of the transient stray field induced by the pump beam ($\lambda$ = 730 nm, $P = 220\;\mu$W, $t_p = 1.2\;\mu$s, $t_d$ = 0, $2\tau = 13\;\mu$s). (c) Left: simulation of the additional stray magnetic field $\Delta B_{NV}(\bold{r}) \propto -\Phi(\bold{r})$ due to a Gaussian enhancement of the in-plane magnetization $\Delta M_x(\bold{r})$. Right: Spatial distribution and linecut of the simulated $\Delta M_x(\bold{r})$ with the black circle and double arrow denoting the full width at half maximums. (d) Maximal NV precession angle $\Phi$ on a 5-layer \CrCl{} region after a single pump pulse of variable fluence. The fluence is varied by changing either the duration $t_p$ or power $P$ of the pump pulse, with the data collapsing onto a single saturation curve.}
\end{figure*}

Using only the green NV probe beam, we switch to a Hahn echo sequence for sensing dynamic fields. Here, the NV center is prepared in a superposition state and allowed to precess over a free evolution time 2$\tau$, separated into two halves by a $\pi$-pulse (Fig. \ref{fig:2}b). The total precession angle $\Phi$ is determined from $\Phi = \arctan(Y_P/X_P)$, where $X_P$ and $Y_P$ are the $X$- and $Y$-axis Bloch sphere projections of the final state (Appendix B). If the stray magnetic field is static over the $2\tau$ evolution, the precession during the first half exactly cancels that over the second half, and $\Phi = 0$. This trivial behavior is indeed observed over the bare NV substrate (0-layer), such that $Y_P$ remains fixed at 0 and $X_P$ decays smoothly due to quantum decoherence as the evolution time is lengthened (Fig. \ref{fig:2}c). Unexpectedly, when we perform the Hahn echo over the 5-layer region, $Y_P$ displays a pronounced increase before being bounded by the decoherence envelope, revealing that the precession angle $\Phi$ increases with evolution time (Fig. \ref{fig:2}d). The same experiment on the 6-layer region, however, exhibits only faster decoherence compared to the bare substrate due to magnetic fluctuations in \CrCl, but negligible coherent phase accumulation (Fig. \ref{fig:2}e).


By scanning the green beam over the diamond and \CrCl, we spatially image the accumulated $\Phi$ for a Hahn echo sequence with fixed evolution time $2\tau = 13\;\mu$s (Fig. \ref{fig:2}f). We observe nearly homogeneous phase precession over the 5-layer region, but negligible precession over the 6 and 8-layer regions. For particular experimental conditions (e.g., coherence time), this imaging scheme can be significantly faster than ODMR for distinguishing even and odd layers due its strong signal in the flake's interior, independent of the flake's dimensions, and fewer data points needed per pixel (Appendix B).

Our results clearly indicate that a time-varying magnetization $\Delta M(t)$ is stimulated in the \CrCl{} flake (e.g., a decaying exponential as diagrammed in Fig. \ref{fig:2}b), such that its stray field is not constant over the two halves of the Hahn echo. One reasonable, but inconsistent, hypothesis is that NV center probe laser locally heats the \CrCl{} flake, causing a transient demagnetization that gradually recovers during the intervening laser off period. If thermal demagnetization were the mechanism, we would expect $\Phi$ to saturate when 2$\tau$ significantly exceeds the thermal time constant ($\approx$1 $\mu$s) (Supplementary Section V) \cite{Zhou2019}. However, the experimental $\Phi$ is growing over tens of microseconds (Fig. \ref{fig:2}d), indicating a much longer magnetization recovery time. Moreover, we will show that the spatial distribution of the transient stray field corresponds to an \emph{enhancement}, rather than a reduction, of the in-plane magnetization.


\subsection{Pump-Probe Imaging of Locally Enhanced Magnetization}
By incorporating a second laser dedicated to pumping \CrCl{}, we can scan the NV probe beam while holding the \CrCl{} excitation fixed to map the stray field profile of the optically-induced magnetization. We pulse the pump laser after the NV probe laser has been turned off for at least 50 $\mu$s to allow the incidental effect of the probe to decay (Fig. \ref{fig:3}a). The polarization of the pump laser is linear and does not alter the results. Crucially, the Hahn echo is still required for sensitivity, as dc sensing with the pump laser on cannot resolve the small magnetization difference.

Figure \ref{fig:3}b shows images of the transient stray field as a function of temperature upon exciting the 7-layer stripe of Fig. \ref{fig:1}a with a pump wavelength $\lambda$ = 730~nm. The spatially-resolved $\Phi(\bold{r})$, where $\bold{r}$ denotes the position of the probe beam relative to the pump, displays a double-lobe, sign-switching feature, which remains virtually unchanged up to a few Kelvin below $T_N$, but disappears at 18 K (above $T_N$). Notably, the positive lobe (red) in $\Phi(\bold{r})$ includes the center point where the pump and probe beams coincide (yellow circle), consistent with the sign for $\Phi$ measured when the single NV beam serves as both pump and probe (Fig. \ref{fig:2}f).

At $B_{ext}$ = 141.5 mT, the phase accumulated by our sensing state during the Hahn echo is given by an integral of the transient stray field $\Delta B_{NV}(t)$:
\begin{equation}\label{eq:1}
	\Phi(2\tau,t_d) = 2\pi \gamma_e \Biggl(\int \limits_{t_d+\tau}^{t_d+2\tau}  \Delta B_{NV}(t') dt' - \int \limits_{t_d}^{t_d+\tau}  \Delta B_{NV}(t') dt'\Biggr).
\end{equation}
Measurements using a simulated waveform for $\Delta B_{NV}(t)$ corroborate that positive $\Phi$ corresponds to $\Delta B_{NV}$ that is more negative (antiparallel to the NV axis) in the first half of the Hahn echo sequence than in the second half (Supplementary Section V).

The localized pattern for $\Phi(\bold{r})$ (Fig. \ref{fig:3}b) motivates us to consider a Gaussian profile, following the pump beam intensity, for the optically-induced magnetization $\Delta M(\bold{r},t)$. In Fig. \ref{fig:3}c, we simulate the additional stray field $\Delta B_{NV}(\bold{r})$ produced by $\Delta M(\bold{r})$ corresponding to an enhancement of the in-plane magnetization along the $+\hat{x}$ direction with Gaussian amplitude and full-width-half-maximum of 1.0~$\mu$m (black circle), approximately the pump beam size. The time-integrated $\Phi(\bold{r})$ measured by Hahn echo should then display the same spatial distribution as $-\Delta B_{NV}(\bold{r})$ since $\Delta B_{NV}(\bold{r},t)$ decays towards zero during the echo duration (Eq. \ref{eq:1}). Fig.~\ref{fig:3}d displays the typical dependence of the maximal $\Phi$ on the pump fluence, showing it to saturate for high powers $P$ or long pulse widths $t_p$, hinting at a metastable state whose population can saturate.

The excellent correspondence between the experimental (Fig. \ref{fig:3}b) and simulated patterns (Fig. \ref{fig:3}c) demonstrates that below $T_N$, the existing in-plane magnetization is locally enhanced by optical excitation. Alternatively, we can allow the Gaussian enhancement $\Delta M(\bold{r})$ to be oriented along an arbitrary angle $\theta_m$ in the $xz$-plane (Supplementary Fig. 3). In this analysis, $\theta_m \approx 0^\circ$ ($+\hat{x}$) still produces the best fit to the experimental $\Phi(\bold{r})$ image. To explain the even-odd contrast in the observable $\Delta B_{NV}$ (Fig. \ref{fig:2}f), we plausibly suppose that the enhancement $\Delta M$ in each \CrCl{} layer, if any, occurs along the magnetization direction of that layer. Then $\Delta B_{NV}$ would be canceled over even layer areas due to AF interlayer coupling, although the underlying effect occurs everywhere.

As $T_N$ is approached upon warming to 16 K, the $\Phi(\bold{r})$ pattern nearly inverts for this specific flake (Fig. \ref{fig:3}b). A higher resolution temperature sweep for a different 7-layer flake, presented in Supplementary Fig. 5, shows a rapid series of changes in $\Phi(\bold{r})$, including progressive rotations of its symmetry axis, near $T_N$. The symmetry axis of $\Phi(\bold{r})$ reflects the vector direction of the transient change in magnetization. For example, we can deliberately rotate $\Phi(\bold{r})$ by adding a vertical ($\hat{y}$) component to the bias field $B_{ext}$ (Supplementary Fig. 6), which rotates the static magnetization of \CrCl{} given its negligible intrinsic anisotropy within the easy plane. Hence, the fickle patterns in $\Phi(\bold{r})$ near criticality could relate to local fluctuations in anisotropy, or to other potential subtleties such as laser heating (demagnetization) or inhomogeneous N\'eel temperature $T_{N}$ \cite{Wahab2020}.




\subsection{Role of Strongly Bound Excitons}
\begin{figure}[t]
\includegraphics[scale=1]{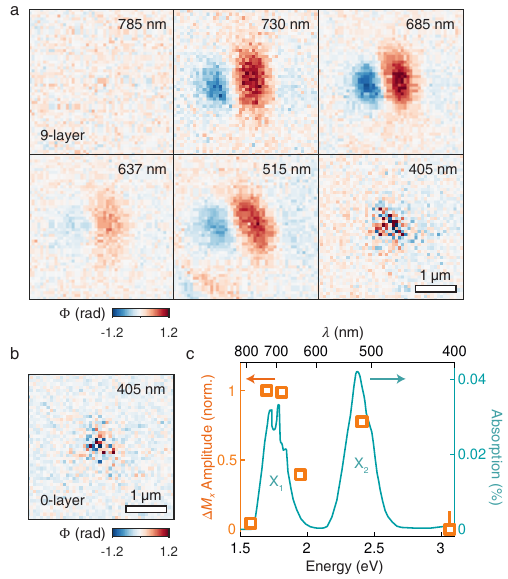}
\caption{\label{fig:4}Dependence of $\Delta M_x$ on the pump wavelength $\lambda$. (a) Dual-beam Hahn echo phase maps $\Phi(\bold{r})$ imaged on a 9-layer region for six different $\lambda$ ($t_d =1\;\mu$s, $2 \tau =20\;\mu$s). The optical power $P$ is 180 $\mu$W, except for $\lambda$ = 405~nm, where $P$ = 55 $\mu$W. For $\lambda <$ 637~nm, the pump beam can excite or ionize the NV center, but this introduces only incoherent photon noise, as visible in the $\lambda =$ 405~nm image. (b) $\Phi(\bold{r})$ imaged on bare diamond with pump $\lambda =$ 405~nm, showing that the speckle noise is due to NV ionization and independent of \CrCl. (c) Orange data points: extracted magnetization enhancement amplitude $\Delta M_x$ (normalized by $\lambda = 730$ nm and for power $P$) versus the pump wavelength $\lambda$. The overlaid teal curve denotes the expected optical absorption for a 9-layer flake, revealing two exciton peaks $X_1$ and $X_2$ (absorption data derived from Ref.~\cite{Bermudez1979}).}
\end{figure}

We focus our attention to understanding the robust magnetization enhancement below $T_N$ by studying its dependence on the \CrCl{} excitation wavelength. Fig.~\ref{fig:4}a shows $\Phi(\bold{r})$ imaged on a 9-layer flake at six different pump wavelengths $\lambda$ between 405~nm and 785 nm. These wavelengths all lie below the ligand-metal charge-transfer absorption edge for \CrCl{} at 3.2 eV (388 nm) \cite{Pollini1970}. To mitigate photon noise associated with excitation of the NV center for $\lambda < 637$~nm (e.g. Fig. \ref{fig:4}b), we use a nonzero delay $t_d = 1\;\mu s$ for all $\lambda$, which allows some fraction of the excited NV centers to relax prior to sensing. This NV charge noise cannot bias the pattern for $\Phi(\bold{r})$, since the coherent signal is contributed only by NV centers present in the correct, microwave-addressable ground state at the start of the Hahn echo. For each $\lambda$, we fit $\Phi(\bold{r})$ according to the Gaussian model (Fig. \ref{fig:3}c) with the beam width and amplitude $\Delta M_x$ of the magnetization enhancement along the $\hat{x}$-direction as free parameters. Figure \ref{fig:4}c shows the extracted $\Delta M_x$ as a function of $\lambda$, normalized by the highest enhancement at $\lambda$ = 730 nm. In the same figure, we overlay the optical absorption spectrum of \CrCl{} calculated from data at 4.2 K in Ref.~\cite{Bermudez1979}, showing it to mimic the wavelength dependence of the optical magnetization enhancement.


The prominent absorption peaks at $\sim$1.7 eV and $\sim$2.4~eV correspond to the two lowest energy bright excitons ($X_1$ and $X_2$) identified in recent many-body perturbation theory calculations of \CrCl{} excited states \cite{Zhu2020b,Acharya2022a}. These excitons derive from the interconfigurational $d$\mbox{-}$d$ transitions of the Cr$^{3+}$ ion ($3d^3$) in an octahedral ligand field that become parity-allowed due to metal-ligand hybridization. The latter $dp$-hybridization increases with halogen size; hence, excitons in \CrCl{} are expected to be the darkest and most tightly bound, while they are the brightest and most delocalized in \CrI{} \cite{Seyler2018,Acharya2022a}.

\begin{figure*}[hbt]
\includegraphics[scale=1]{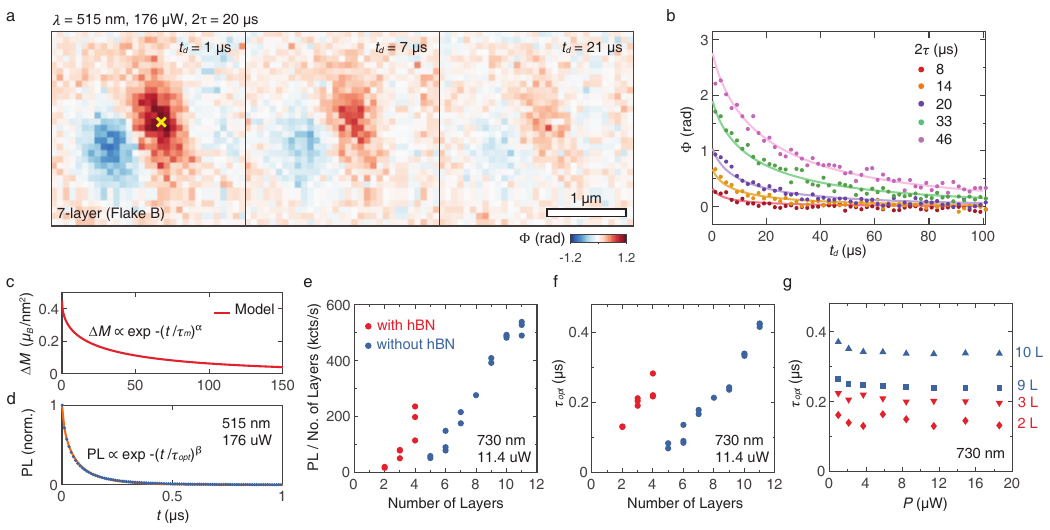}
\caption{\label{fig:5}Time-resolved magnetization and PL measurements. (a) Spatiotemporal images of $\Phi(\bold{r})$ for various delays $t_d$ between the pump pulse and Hahn echo sequence on 7-layer \CrCl{}. The rotation of the $\Phi(\bold{r})$ symmetry axis away from the $+x$-direction here is atypical for 4 K and likely reflects a local in-plane anisotropy that competes with the external field $B_{ext}$. (b) Hahn echo phase $\Phi$ measured at the yellow cross in a) as a function of the pump-probe delay $t_d$ and total echo duration $2\tau$. The solid lines are simultaneous best fit curves using a magnetization decay model $\Delta M(t) = \Delta M_0 \exp(-(t/\tau_m)^\alpha)$. (c) Inferred magnetization decay from b), yielding $\Delta M_0 = 0.46\;\mu_B$/nm$^2$, $\tau_m = 26\;\mu s$, and $\alpha = 0.5$. (d) Decay of the PL intensity for 7-layer \CrCl{} after excitation with $\lambda$ = 515 nm ($X_2$). The solid best fit line extracts a long exciton lifetime $\tau_{opt} = 38$ ns and exponent $\beta = 0.6$. (e) PL intensity per layer versus number of \CrCl{} layers for $\lambda$ = 730 nm excitation ($X_1$). (f) Exciton lifetime $\tau_{opt}$ versus number of \CrCl{} layers for  $\lambda$ = 730 nm, with $\beta = 0.79$ fixed for all. (g) Dependence of $\tau_{opt}$ on the excitation power $P$ for selected layer thicknesses. The red (blue) points in e), f), g) denote data from regions encapsulated (not encapuslated) by hBN.}
\end{figure*}

To clarify the role of excitons, we compare the lifetime of the optically-induced magnetization $\tau_m$ to the exciton lifetime $\tau_{opt}$. In Fig. \ref{fig:5}a, we visualize the decay of $\Phi(\bold{r})$ for increasing delay $t_d$ between the pump pulse and Hahn echo sequence on a 7-layer region. To quantitatively extract $\tau_m$, we park the probe at the position of highest intensity in $\Phi(\bold{r})$ (yellow cross in Fig. \ref{fig:5}a) and measure the acquired $\Phi$ as a function of both the delay $t_d$ and total evolution time $2\tau$ (Fig. \ref{fig:5}b). We then use Eq. \ref{eq:1} to simultaneously fit this set of curves, assuming a phenomenological form of the magnetization decay $\Delta M(t) = \Delta M_0 \exp(-(t/\tau_m)^\alpha)$ that creates $\Delta B(t)$ through the Gaussian model (Supplementary Section VI). The best-fit $\Delta M(t)$ curve is plotted in Fig. \ref{fig:5}c, displaying an estimated magnetization lifetime $\tau_m = 26 \pm 3\;\mu s$ with exponent $\alpha = 0.5 \pm 0.1$ and a maximal magnetization change $\Delta M_0 = 0.46 \pm 0.1\;\mu_B$/nm$^2$.


In contrast, the time-resolved PL intensity for a representative 7-layer \CrCl{} flake (Fig. \ref{fig:5}d), isolated with a 800 nm long-pass filter \cite{Cai2019a}, completely decays within 1 $\mu$s, indicating an exciton lifetime $\tau_{opt} \approx 40$ ns for 515 nm excitation ($X_2$). Since $\tau_{opt}$ is two orders of magnitude shorter than $\tau_m$, the exciton itself cannot be the magnetically enhanced state, but rather must give rise to the latter during its recombination as a long-lived intermediate state.

As excitons in \CrCl{} are nearly radiatively forbidden, we anticipate that nonradiative recombination could present the primary limit to $\tau_{opt}$. In Fig. \ref{fig:5}e, we display the PL intensity normalized by the number of layers at 730~nm excitation ($X_1$) for several \CrCl{} regions between 2 to 11 layers thick. The red (blue) data points denote regions encapsulated (not encapsulated) by hBN, where both samples were prepared in the glovebox and simultaneously loaded into the cryostat. Although we observe variations among regions of identical thickness, the PL \emph{per layer} for both conditions decreases for thinner flakes, a trend similar to \CrI{} \cite{Seyler2018}. Simultaneously, the exciton lifetime $\tau_{opt}$ is reduced with decreasing thickness (Fig. \ref{fig:5}f), with flakes lacking hBN encapsulation having shorter lifetimes and lower PL for comparable thickness. Taken together, Figs. \ref{fig:5}e,f clearly identify the rise of nonradiative transitions, as opposed to diminishing radiative oscillator strength, as the primary mechanism for PL quenching in thinner layers.



Our evidence that exciton pumping creates an alternative magnetically enhanced electronic state hints that the energy dissipated in nonradiative exciton recombination may be transferred to a nearby electronic carrier, a channel known as Auger recombination. Moreover, the sensitivity of $\tau_{opt}$ to layer thickness and surface encapsulation (Fig. \ref{fig:5}f) implies that Auger recombination in \CrCl{} is catalyzed by defects residing in the surface layers. Excitons produced in the bulk can hop between layers \cite{Ginsberg2020} within a finite diffusion length to reach the surface, where the recombination rate is highest. The rate for defect-assisted Auger recombination crucially depends on the probability of finding the exciton's electron and hole near each other, and hence should be strongly enhanced in \CrCl{} with its near-atomic exciton radii \cite{Chen2001, Wang2015g}. The conventional Auger process for delocalized Wannier-Mott excitons in narrow bandgap semiconductors is instead governed by exciton-exciton collisions, resulting in an exciton lifetime that is inversely proportional to the square of the exciton density in the high-density limit \cite{McAllister2018}. In contrast, Fig. \ref{fig:5}g shows that the lifetime $\tau_{opt}$ in few-layer \CrCl{} is independent of the optical excitation power, assumed to be proportional to the exciton density. This independence from the exciton density is a direct signature of defect-assisted Auger recombination of single localized excitons \cite{Chen2001, Wang2015g}.



\subsection{Controllable, Surface-Sensitive Magnetization Enhancement}
\begin{figure}[t]
\includegraphics[scale=1]{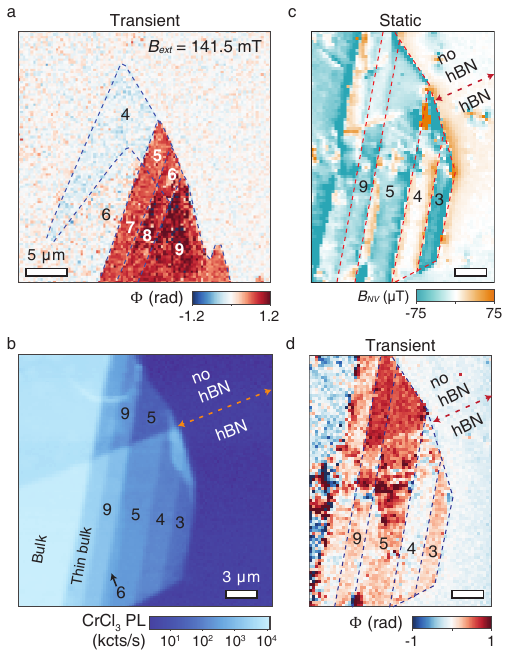}
\caption{\label{fig:6}Surface-sensitive magnetization enhancement. (a) Hahn image for the 4-9 layer \CrCl{} flake (hBN encapsulated) of Fig. \ref{fig:1} at high $B_{ext}$ = 141.5 mT. The Hahn precession angle $\Phi$ with $2\tau = 13\;\mu$s is measured as the single green beam (515 nm) is rastered over the flake. The amplitude of the transient field over the 6/8-layer stripe is similar to the neighboring odd-layer regions even though the former possesses a twice larger volume magnetization. (b) PL image ($>$800 nm) for a 3-9 layer \CrCl{} flake with partial hBN encapsulation. The edge of the hBN cuts across the entire flake parallel to the dashed orange line. The PL is quenched over unencapsulated regions due to defect-assisted, nonradiative exciton recombination. (c) Static stray field ($B_{NV}$) image of the same flake confirming even-odd layer parity and displaying slightly reduced stray fields over the unencapsulated area. (d) Hahn image of the same flake with $2\tau = 19\;\mu$s. The transient stray field, proportional to $\Phi$, strongly increases in the exposed areas of the odd-layer stripes, indicating that surface impurities catalyze the optical magnetization enhancement.}
\end{figure}

We next demonstrate that the optically-induced magnetization is surface-dominated and sensitive to surface condition. First, for the 6/8-layer stripe of Fig. \ref{fig:1} at high $B_{ext}$, we observe a dc stray field that is approximately twice as strong as its neighboring odd-layer regions (Fig. \ref{fig:1}e), indicating an extra flipped layer in its volume. However, its transient stray field under exciton pumping is virtually identical in strength to the nearby odd-layer regions (Hahn map, Fig. \ref{fig:6}a). This is explained if the surface monolayers dominate the contribution to the optical enhancement since if all layers contributed equally, the transient signal would also be twice as strong. Indeed, the flipped layer in the 6/8-layer stripe is expected to be a surface layer, since this configuration minimizes the energy penalty for breaking AF interlayer exchange. 

Second, we fabricate another \CrCl{} flake where we deliberately encapsulate only a portion of the flake with hBN. This allows an unequivocal comparison of the surface condition by ensuring that the same starting sample experiences identical exposures inside the glovebox and in ambient during sample loading. In Fig. \ref{fig:6}b, we plot the PL of the \CrCl{} flake on diamond using a logarithmic colorbar to emphasize thinner regions. Areas of the same layer thickness are significantly dimmer when not encapsulated by hBN, corroborating the role of surface defects in promoting nonradiative exciton recombination (Fig. \ref{fig:5}e,f.).

Figure \ref{fig:6}c displays the static stray field ($B_{NV}$) image of the same flake, showing that antiferromagnetic interlayer coupling is retained down to three layers. Strikingly, the transient stray field image obtained under exciton pumping with the single green beam displays a prominent contrast between the exposed and covered regions (Fig. \ref{fig:6}d). The average Hahn echo phase $\Phi$ recorded over the exposed portion of the odd-layer stripes increases by more than a factor of two over the hBN covered portion ($\Phi_{exposed} \approx 0.60$ rad, $\Phi_{covered} \approx 0.25$ rad), excluding a transitional region near the termination of the hBN. This comparison pinpoints environmentally-induced defects in the surface layers as the common catalyst for optically-enhanced magnetization (Fig. \ref{fig:6}d) and nonradiative exciton recombination (Fig. \ref{fig:6}b).


\begin{figure*}[hbt]
\includegraphics[scale=1]{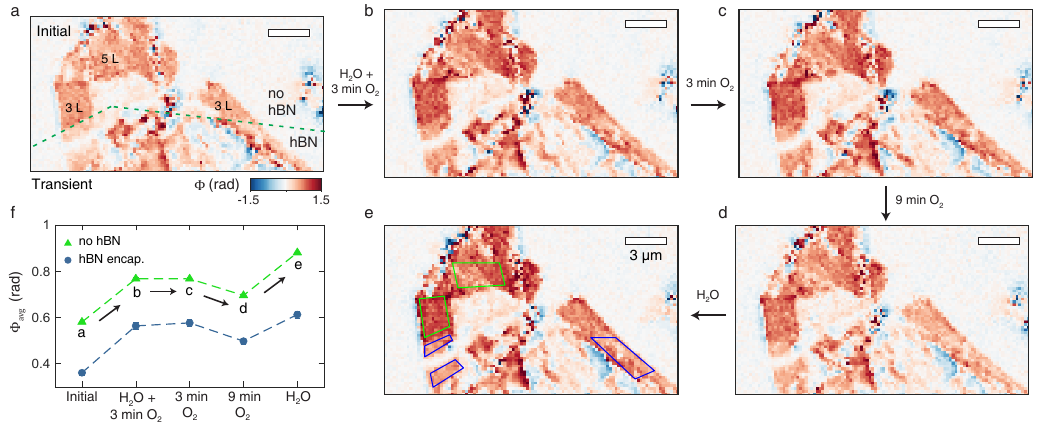}
\caption{\label{fig:7}Magnetization enhancement under controlled gas exposure. (a) Initial $\Phi$ image (single beam, $\lambda  = 515$ nm, $P$ = 50~$\mu$W, $2\tau = 19\;\mu$s) for an as transferred flake, containing 3 and 5-layer regions. The edge of the hBN encapsulation is delineated by the green dashed line.  (b) $\Phi$ image after thermal cycling the sample without pumping (equivalent to H$_2$O exposure at $5\cdot10^{-2}$~mBarr for approximately 2 hours) and then exposing it to 200 mBar pure O$_2$ gas for 3 minutes. (c) Additional exposure to only O$_2$ (200 mBar, 3 minutes). Here, the sample was thermal cycled under continuous pumping to limit H$_2$O pressures below $1\cdot10^{-4}$ mBarr. (d) Additional exposure to only O$_2$ (200 mBarr, 9 minutes). (e) Additional exposure to only H$_2$O (thermal cycling without pumping). (e) Summary of the evolution of $\Phi$ versus sequential surface treatment. The green (blue) data denotes $\Phi$ over the exposed (encapsulated) regions of the flake as determined by the average over the regions defined by the matching colored boxes shown in (e). The transient magnetization enhancement, proportional to $\Phi$, increases after exposure to H$_2$O, but not O$_2$.}
\end{figure*}

To probe the origin of the surface defect responsible for these simultaneous effects, we expose a few-layer \CrCl{} flake to controlled doses of water vapor or oxygen gas, the two main reactive species residual in our glovebox and in ambient. Fig. \ref{fig:7}a presents the initial transient field image $\Phi(\bold{r})$ for the sample, again partially-covered by hBN. A baseline level for $\Phi$ is measured despite minimized sample preparation (3 hours in glovebox) and transfer time (110 seconds in ambient). After thermal cycling the sample without pumping, we then deliberately expose the surface to 200 mBarr of O$_2$ gas for 3 minutes at room temperature (290 K). This treatment increases $\Phi$ by $\approx$0.2 radians in the both the exposed and encapuslated regions (Fig. \ref{fig:7}b), indicating that single-sided encapsulation does not form a gas-tight seal \cite{Holler2020}. When our cryostat is warmed up without pumping, a pressure of approximately $5\cdot10^{-2}$ mBarr is measured after crossing 273 K, which is dominated by outgassed H$_2$O molecules. Residual gas analysis indicates the ratio of H$_2$O to O$_2$ in this state is 160:1. Hence, although this first iteration demonstrates control over the surface defect density and the amplitude of the transient magnetization, it does not yet distinguish between H$_2$O or O$_2$ exposure as the cause.

We perform a second thermal cycle where we instead continuously pump on the cryostat during warming, which limits the evolved H$_2$O  pressure to $<1\cdot10^{-4}$ mBarr (brief spike near 273 K). We then quickly dose the surface again with 200 mBarr of O$_2$ for 3 minutes (Fig.~\ref{fig:7}c). In this case, the sample experiences only exposure to O$_2$, and no change in $\Phi$ is observed. To confirm this result, we repeat the thermal cycle under pumping and expose the sample to an additional 9 minutes of O$_2$ (Fig. \ref{fig:7}d). The measured transient field $\Phi$ actually slightly decreases after this treatment, indicating that O$_2$ exposure may displace the actual active surface impurity. In contrast, exposing \CrCl{} to only H$_2$O by a final thermal cycle without pumping and with no introduction of O$_2$ increases $\Phi$ once more (Fig.~\ref{fig:7}e). The evolution of $\Phi$ after sequential surface treatments is summarized in Fig. \ref{fig:7}f and definitively proves that H$_2$O exposure, rather than O$_2$, provides the dominant catalyst for our effect.

\subsection{Defect-Assisted Exciton Recombination Mechanism}
\begin{figure}[thb]
\includegraphics[scale=1]{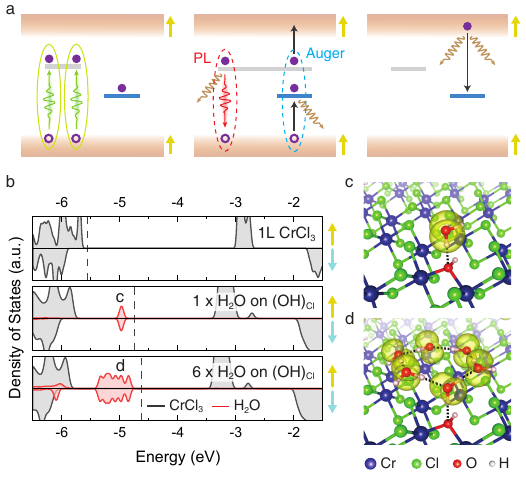}
\caption{\label{fig:8}Proposed mechanism for magnetization enhancement through defect-assisted Auger recombination. (a) Mechanism. Left: near-resonant excitation of \CrCl{} creates strongly bound excitons (filled circle - electron, open circle - hole). An in-gap defect state is occupied. Center: Excitons can recombine radiatively, resulting in PL (red), or nonradiatively through defect-assisted Auger recombination (blue). The three-particle Auger process converts, in effect, an unpolarized defect electron into a spin-up \CrCl{} conduction electron. Phonons (brown arrows) can participate in both radiative and Auger recombination. Right: the excited state after Auger recombination decays by multi-phonon emission and electron capture by the ionized defect. (b) Spin-resolved density of states for a pristine \CrCl{} monolayer (top); a monolayer with a single H$_2$O molecule adsorbed onto (OH)$_\mathrm{Cl}$, denoting an $-$OH group that substitutes for a Cl atom (middle); a monolayer with six H$_2$O molecules adsorbed onto (OH)$_\mathrm{Cl}$ (bottom). Gray (red) lines indicate the contributions from monolayer \CrCl{} (H$_2$O). The Fermi level is shown as the vertical dashed line. (c,d) Spatial spin distributions, doubly degenerate, for the localized, in-gap states labeled in (b), contributed by a single H$_2$O adsorbate and a cluster of six H$_2$O adsorbates, respectively. The dashed lines denote hydrogen bonds.}
\end{figure}


The experimental signatures thus point toward the following defect-assisted Auger recombination mechanism \cite{Chen2001, Wang2015g} to stimulate the optically-enhanced magnetization in atomically thin \CrCl. Auger recombination of excitons is mediated by an H$_2$O-induced surface impurity, which introduces an occupied mid-gap level. During recombination, this defect level captures the hole of the exciton and transfers the excess energy to its electron partner, injecting it into the spin-polarized conduction band (Fig. \ref{fig:8}a, middle). This effectively converts an unpolarized defect electron into a spin-up conduction electron, increasing the system magnetization by $\approx$1$\mu_B$ per event. To conserve total angular momentum, phonons generated by localized vibrations of the defect state must also participate in the recombination \cite{Tauchert2022}. Finally, the resulting excited state persists until the ionized defect level recaptures an electron from the conduction band (Fig. \ref{fig:8}a, right), which is accompanied by multi-phonon emission and results in the slow magnetization recovery time $\tau_m$ observed at cryogenic temperatures.

For monolayer \CrI{}, electron doping via halogen vacancies \cite{Zhao2018,Pizzochero2020} or metal adsorption \cite{Guo2018a,Yang2021} has indeed been proposed to increase the electron charge (from $3d^3$ to $3d^4$) and magnetic moment of the Cr ion, since a high spin configuration is preferred for the weak Cl ligand field. In the band picture, this amounts to partial filling of the lowest energy, majority-spin conduction bands contributed primarily by the Cr $e_g$ orbitals. In our mechanism, we realize this doping effect in the surface monolayers of \CrCl{} through optically inducing charge transfer from surface impurities. Since the surface monolayers of \CrCl{} dominate the magnetization enhancement, an even-odd effect is naturally produced: in odd-layer (even-layer) regions, the top and bottom layers have parallel (anti-parallel) magnetizations. Assuming that $\Delta M_0 \approx 0.5\;\mu_B$/nm$^2$ is contributed equally by the two surfaces, an excess electron density of $2.5\cdot10^{13}/$cm$^2$ is induced locally inside the beam spot in each surface monolayer, which is comparable to levels achieved by electrostatic doping of \CrI{} \cite{Jiang2018a}. In the latter experiment, complementary signatures to enhanced magnetization, such as enhanced coercive fields and critical temperatures, were also observed \cite{Jiang2018a}.


To quantitatively evaluate the plausibility of defect-assisted Auger recombination, we turn to DFT calculations. For a pristine \CrCl{} surface, we find that the adsorption energy for an H$_2$O molecule ($E_{ads} = -0.15$~eV) is greater than the Gibbs free energy of gas-phase H$_2$O at ambient conditions ($E_{gas} \approx -0.5$~eV) \cite{Lemmon2018}; hence, direct physisorbed H$_2$O should be thermodynamically unfavorable (Supplementary Sec. VIII). However, Cl vacancies are known to be prevalent in \CrCl{} \cite{Mastrippolito2021}. The excess electron density near this vacancy may provide a natural site for hydrolysis and the binding of an $-$OH group, forming a substitutional (OH)$_\mathrm{Cl}$ defect. Subsequent H$_2$O molecules will form hydrogen bonds to the randomly distributed $-$OH groups and thereafter to each other for greater stability. As a concrete, but arbitrary example, a ring configuration of six H$_2$O molecules hydrogen-bonded around a single (OH)$_\mathrm{Cl}$ (Fig. \ref{fig:8}d) is found to possess $E_{ads} = -0.47$~eV per molecule. Thus, Cl vacancies provide attachment points to nucleate the growth an H$_2$O absorbate layer, whose presence has been deduced from surface science experiments \cite{Liu2019,Paolucci2024}.


Fig. \ref{fig:8}a (top) presents our DFT calculation of the spin-resolved, projected density of states for \CrCl, displaying majority-spin valence and conduction bands near the Fermi level. A single H$_2$O molecule adsorbed at (OH)$_\mathrm{Cl}$ introduces a spin-degenerate defect state inside the \CrCl{} band gap (Fig. \ref{fig:8}b, middle). The spin-up density for this state (identical to spin-down) is depicted in Fig.~\ref{fig:8}c, showing it to derive primarily from the nonbonding oxygen $p$-orbital that is perpendicular to the H$_2$O molecule plane. For the six H$_2$O cluster, the in-gap state broadens into a defect band due to orbital hybridization (Fig. \ref{fig:8}b, bottom). These in-gap levels should appear generically for a disordered, weakly bonded H$_2$O layer and could plausibly mediate Auger recombination of excitons. They are theoretically predicted to be energetically situated closer to the valence band for efficient hole capture, while also being within $\sim$1.7 eV of the conduction band to allow the $X_1$ exciton to scatter an electron into it. Alternatively, we find that O$_2$-related defects and intrinsic vacancy defects do not provide appropriate conditions for optical magnetization enhancement (Supplementary Sec. VIII). The former possess large formation energies, making them thermodynamically unfavorable. The latter vacancy defects introduce spin-majority in-gap states that already increase the magnetization of \CrCl{} without optical excitation, and moreover, their density should be unchanged by our gentle gas exposures.

\subsection{Conclusions}
We have revealed that exciton pumping enhances the in-plane, layer-alternating magnetization of atomically-thin \CrCl{}. Experimental and theoretical evidences support a mechanism where the Auger recombination of strongly bound excitons in the surface layers activates charge transfer between water-induced adsorbates and the Cr majority-spin bands. Through an innovative quantum coherent sensing protocol, our work demonstrates that impurities can enhance the intrinsic magnetism in single atomic layers \cite{Zhao2018,Guo2018a,Pizzochero2020,Yang2021,Wang2018d,Luo2020}. Exciton-mediated charge transfer may be applicable to diverse semiconducting 2D magnets, including intrinsic Cr- and Mn-based compounds and magnetically-doped TMDs, where long-range order in the latter are still under debate. These systems, including CrX$_3$ \cite{Wu2019a,Molina-Sanchez2020,Wu2022,Zhu2020b,Acharya2022a}, CrPS$_4$ \cite{Kim2022}, MnPX$_3$ \cite{Birowska2021}, Fe-doped MoS$_2$ \cite{Fu2020} and V-doped WSe$_2$ monolayers \cite{Nguyen2021a}, all harbor tightly bound excitons, either due to nearly forbidden $d$-$d$ transitions or to reduced dielectric screening, and are susceptible to deep-level states. Indeed, our study opens strategies for future efforts to optimize the magnetization enhancement by extending to bulk layers via doping, intercalation, or defect creation through irradiation.

Optical control is furthermore attractive as a high-bandwidth knob to modulate the resonant coupling between 2D spin ensembles and small mode volume superconducting resonators for solid-state quantum memory applications \cite{Zollitsch2023,Kubo2011}. Conventionally achieved by pulsing an external magnetic flux, tuning the magnon resonances and microwave resonator in and out of match could alternatively be realized with fast and focusable pulses of light that change the magnetization and internal field of the 2D spin ensemble \cite{Macneill2019,Singamaneni2020}. Finally, further development of the introduced pump-probe quantum magnetometry technique to extend its sensitivity and spatiotemporal resolution could open a new window into light-induced transient magnetism, superconductivity \cite{Fava2024}, and charge dynamics in condensed matter systems.


%

\section{Appendix A: Sample Details}
The experimental results have been verified on multiple \CrCl{} crystals grown by two different academic groups and by a commercial vendor (2D Semiconductors; Fig. \ref{fig:5}).  For Figs. \ref{fig:1}-\ref{fig:4}, the \CrCl{} crystals were synthesized by vacuum sublimation of chromium (III) chloride powder (99.9\%) in a single-zone tube furnace. The powder material with a total mass of 500 mg was sealed inside an evacuated silica tube, which was transferred into the furnace and heated to 650 $^\circ$C at 5 $^\circ$C/min, kept at that temperature for 72 hours, and cooled to room temperature. Large crystals of CrCl$_3$ with millimeter-size lateral dimensions and micron-size thickness was obtained. For Figs. \ref{fig:6} and \ref{fig:7}, the \CrCl{} crystals was synthesized by vacuum sublimation of chromium (III) chloride powder (99.9\%, Strem, USA) in a two zone  tube furnace in a quartz ampoule. 25 g of \CrCl{} were placed in an ampoule (50x250 mm) and melt sealed under high vacuum of diffusion pump with liquid nitrogen cold trap at pressure under 1x10$^{-3}$ Pa. The sealed ampoule was placed in two zone horizontal furnace. First the growth zone was heated on 700 $^\circ$C, while the source zone was kept at 500 $^\circ$C. After two days the thermal gradient was reversed and source zone was kept at 700 $^\circ$C while the growth zone temperature was slowly reduced from 680 to 600 $^\circ$C over a period of one week and for additional one week was kept at 600 $^\circ$C. Over 90\% of the material was transported forming plate crystals with size up to 30 mm and thickness under 0.1 mm.

Two diamond substrates are used in this work, both containing 200$\pm$133 nm of $^{12}$C isotopically-purified diamond ($>$99.995\% $^{12}$C as per secondary ion mass spectrometry characterization). Pre-growth, these electronic grade bulk diamond substrates (Element Six) were fine-polished (Syntek) down to $\leq$0.3 nm root mean square (RMS) surface roughness. Thereafter, they underwent a cycling inductively coupled plasma–reactive ion etch (ICP-RIE) to remove the polish-induced damage (five cycles of Ar 25 sccm, Cl$_2$ 40 sccm, 10 mTorr, 400 W ICP, 200 W bias and O$_2$ 50 sccm, 10 mTorr, 700 W ICP, 100 W bias). The ICP-RIE process resulted in $\approx$2.5 $\mu$m removal while maintaining $\leq0.3$ nm RMS. This was followed by a multi-step annealing process at $<$2$\cdot10^{-6}$~Torr (1.6~$^\circ$C/min ramp, 12 h at 200~$^\circ$C, 8 h at 400~$^\circ$C, 8 h at 850 $^\circ$C, and 2 h at 1200 $^\circ$C). Overgrowth was performed in a custom-configured microwave plasma chemical vapor deposition system (SEKI DIAMOND SDS6350) at 11.5~W/mm$^2$, 0.2:400 sccm $^{12}$CH$_4$:H$_2$, and maintained at $\approx$850 $^\circ$C and 25 Torr. Both the hydrogen and the methane were purified to better than 8 and 6 ``nines'', respectively, with the chamber being pumped down to $<$$5\cdot10^{-8}$ Torr base pressure before growth. Before the introduction of the methane precursor, the hydrogen plasma was allowed to run for 20 min as to etch away any residual surface carbonaceous contaminants. 

 
To create a shallow layer of NV centers, one diamond sample was implanted with 40 keV $^{14}$N ions at $10^{12}$~ions/cm$^2$ and annealed at 1050 $^{\circ}$C in forming gas (5\% H$_2$, 95\% Ar) for two hours, while the other was implanted with 45 keV $^{15}$N ions $10^{12}$ ions/cm$^2$ and annealed at 1050 $^{\circ}$C for three hours. At $B_{ext}$ = 141.5 mT, the average Ramsey $T_2^*$ coherence time on the bare diamond is 3 $\mu$s and the average Hahn echo $T_2$ coherence time is 80~$\mu$s. This $T_2^*$ is a factor of ten longer than similar ensemble samples in non-isotopically purified diamond.
 
\CrCl{} and hBN flakes are exfoliated onto separate polydimethylsiloxane (PDMS) stamps using blue tape (Ultron 1009R) inside an argon-filled glovebox with nominal residual concentrations of O$_2$ and H$_2$O $<$0.01 ppm. Air-sensitive atomically thin samples are known to still undergo chemical reactions inside a glovebox, despite impurity gas concentrations beneath the detection limit of standard analyzers \cite{Zalic2019}. A few-layer \CrCl{} flake is then identified by optical contrast (Supplementary Section II). The target \CrCl{} flake is transferred onto the diamond substrate and encapsulated by a suitable hBN flake using sequential PDMS viscoelastic transfer at room temperature. The diamond substrate is taken out of the glovebox, mounted onto a sample holder, and loaded into a closed-cycle optical cryostat (Montana Instruments) for pump-out and cool-down in less than 30 minutes.

\section{Appendix B: NV Center Measurements}
NV magnetic imaging is performed in a custom-built, dual-beam cryogenic (4 K) confocal microscope by pixel-to-pixel scanning of a diffraction-limited probe beam (FWHM = 400 nm). The linearly-polarized pump beam is slightly defocused and has a FWHM around 1 $\mu$m. The two beams are overlapped by analyzing the reflected camera image off the diamond surface. Full details of the optical and electronic setup are provided in Supplementary Section III. Due to the long $T_2^*$ in our isotopically purified diamond samples, we can use pulsed ODMR sequences with a long 800~ns $\pi$-pulse resonant with the $\ket{m_s = 0}\rightarrow\ket{m_s = -1}$ spin transition. Pulsed ODMR resolves the nitrogen hyperfine splitting and enhances the dc field ($B_{NV}$) sensitivity, but requires more data points per pixel (e.g., 25 frequency points to sample the triply-split $^{14}$N resonance lineshape).  For Hahn ($\Phi$) imaging, we acquire four data points per pixel corresponding to four different phases for the final $\pi$/2 projection pulse (30 ns width): $X_{\pm \pi/2}$ and $Y_{\pm \pi/2}$, where the denoted $X$- and $Y$-axis Bloch sphere rotations are realized by in-phase (I) and quadrature (Q) modulation of the microwave drive. The $X_P$ and $Y_P$ projections are then determined differentially by $X_P =  C(X_{-\pi/2}) - C(X_{+\pi/2})$ and $Y_P = C(Y_{-\pi/2}) - C(Y_{+\pi/2})$, where $C$ denotes the integrated photon counts, and then normalized to the range $\pm1$ by the Rabi contrast.


\section{Appendix C: Density Functional Theory Calculations}
Spin-polarized DFT calculations are performed using the Vienna \emph{Ab initio} Simulation Package (VASP) within the projector augmented wave method \cite{Kresse1999}. The electron exchange-correlation is described using the generalized gradient approximation as parameterized by Perdew-Burke-Ernzerhof \cite{Perdew1996}. The van der Waals interaction is accounted for using the DFT+D3 approach \cite{Grimme2010}. The correlation effect is modeled by applying $U_{\mathrm{eff}}$ = 3.5~eV on the Cr $3d$ electrons \cite{Yekta2021} using the DFT+U approach developed by Dudarev \emph{et al.} \cite{Dudarev1998}. The kinetic energy cut-off is set to 400 eV. The first Brillouin zone is sampled by a two-dimensional $\Gamma$-centered $k$-point mesh with $\sim$0.06~\AA{} intervals; for example, an 18×18x1 mesh is used for a monolayer \CrCl{} primitive cell. Electronic energy convergence is achieved with a precision of $10^{-7}$~eV. Lattice and ionic relaxation processes are carried out until the Hellmann-Feynman forces exerted on each ion are reduced below 1~meV/\AA.


\section{Acknowledgments}
The authors thank B. Flebus for valuable discussions. B.B.Z. acknowledges the National Science Foundation (NSF) CAREER award No. DMR-2047214 and NSF award No. ECCS-2041779. This material is based on work supported by the Air Force Office of Scientific Research under award no. FA2386-21-1-4095. B.Y. acknowledges the financial support by the European Research Council (ERC Consolidator Grant ``NonlinearTopo'', No. 815869) and the ISF - Personal Research Grant (No. 2932/21). The diamond synthesis work was primarily supported by the U.S. Department of Energy, Office of Basic Energy Sciences, Materials Science and Engineering Division (N.D., J.A., D.D.A., F.J.H.). F.T. acknowledges support from the U.S. Department of Energy, Office of Basic Energy Sciences, Division of Physical Behavior of Materials under Award No. DE-SC0023124. Z.S. was supported by ERC-CZ program (project LL2101) from Ministry of Education Youth and Sports (MEYS) and used large infrastructure from project reg. No. CZ.02.1.01/0.0/0.0/15\_003/0000444 financed by the EFRR. K.W. and T.T. acknowledge support from the JSPS KAKENHI (Grant Numbers 21H05233 and 23H02052) and World Premier International Research Center Initiative (WPI), MEXT, Japan. M.J. acknowledges support from IITP grant funded by the Korean government (MSIT) (2021-0-01511). This work was performed, in part, at the Integrated Sciences Cleanroom and Nanofabrication Facility at Boston College. 

\end{document}